\documentclass[11pt]{article}
\usepackage{graphicx}              

\newcommand{\bc}{\begin{center}}
\newcommand{\ec}{\end{center}}

\begin{document}

\newcommand{\bu}{{\bf u}}

\begin{titlepage}
\date{}
\title{{\Large\bf BLACK-HOLE UNCERTAINTY ENTAILS AN INTRINSIC TIME ARROW \\
A Note on the Hawking-Penrose Controversy
}}
\author{\Large Avshalom C. Elitzur$^{a}$, Shahar Dolev$^{b}$
}
\maketitle
\begin{center}
$^{a}$ Chemical Physics Department,\\ Weizmann Institute of Science, \\ 76100
Rehovot, Israel.\\

\noindent
E-mail:cfeli@weizmann.ac.il

\vspace{0.1cm}

$^{b}$  The Kohn Institute for the History and  Philosophy of Sciences,\\
Tel-Aviv University, 69978 Tel-Aviv, Israel.\\

\noindent
E-mail:shahar@email.com

\end{center}
\thispagestyle{empty}
\abstract\noindent
Any theory that states that the basic laws of physics are time-symmetric must
be strictly deterministic. Only determinism enables time reversal of entropy
increase. A contradiction therefore arises between two statements of Hawking. A
simulation of a system under time reversal shows how an intrinsic time arrow
re-emerges, destroying the time reversal, when even slight failure of
determinism occurs. 


\begin{center}
\end{center}
\end{titlepage}


The Second Law of Thermodynamics has been dismissed by the majority of authors
as observer-dependent. As proved by Hawking \cite{Haw}, has the Universe's
entropy increase been reversed, this reversal would be impossible to observe.
This is because the time orientation of all biological processes (as we show
elsewhere in detail \cite{Eli,Dol}) relies solely on entropy's increase.
Consequently, our memory and our distinction between "past" and "future" are
similarly oriented. Has the Universe's entropy increase been reversed, memory
and perception would run backwards too. It therefore cannot be ruled out that
we actually live in a universe whose entropy is decreasing.

One might try to dismiss this possibility by arguing that such unique initial
conditions that lead to entropy decrease are highly improbable. Unfortunately,
probabilistic arguments become circular when applied to the entire Universe
\cite{Eli1}. We assign different probabilities to past and future states, yet
{\it the very notions of "past" and "future" are, by convention, based on the
entropy gradient}. When observing a system's evolution, one assigns the
temporal designations "earlier" and "later" according to the universal entropy
gradient prevailing outside the system. No such external reference arrow exists
for the Universe itself. Probability theory cannot, therefore, rule out that
the Universe's entropy increase, as well as all the related perceptual
processes, run backwards. Rather, within the four-dimensional Minkowskian
spacetime, entropy can be equally described as increasing or decreasing,
depending on one's arbitrary choice of time direction \cite{Pri}.

What, then, is the cause of entropy's "increase"? Hawking \cite{Haw1} and
Penrose \cite{Pen} have long been taking opposing views on this issue, their
debate being recently published as a book \cite{Haw2}. Penrose believes that
the long-desired theory of quantum gravity will eventually reveal an intrinsic
time-asymmetry that would account for the macroscopic entropy increase.
Hawking, in contrast, argues that entropy increase only reflects some unique
initial conditions in the universe's evolution (\cite{Haw2}, p. 8). Causation
itself, he stresses, is perfectly time-symmetric: 

\begin{quote}
So if state A evolved into state B, one could say that A caused B. But one
could equally well look at it in the other direction of time, and say that B
caused A. So causality does not define a direction of time (\cite{Haw3}, p.
346).  
\end{quote}

On one point, however, both adversaries agree. For reasons revealed long ago by
Hawking \cite{Haw4,Haw5}, black holes must eventually evaporate in the form of
purely thermal radiation. Now, both Penrose and Hawking agree that all the
information about the objects that have fallen into the black hole is destroyed
when the black hole evaporates. Unlike the ordinary loss of information due to
mixing or noise (which can, in principle, be retrieved), information loss by
black hole evaporation is absolute. In Hawking's words: "quantum gravity
introduces a new level of unpredictability into physics over and above the
uncertainty usually associated with quantum theory" (\cite{Haw2}, p. 60).

We would like to show that Hawking's two assertions are mutually incompatible.
If information is destroyed, by whatever process, then time's arrow is inherent
to causality itself. This conclusion would then rule out the awkward
possibility that entropy increase, with the concomitant psychological time
arrow, run backwards. 

It should be stressed that we do not attempt to prove Hawking's
information-loss hypothesis but only point out its surprising consequences. Our
conclusion, however, rigorously applies to any other theory that makes a
similar assumption, be it the GRW model of spontaneous collapse \cite{Ghi},
Penrose's \cite{Pen} hypothesis concerning the role of gravity in quantum
measurement, or any other assertion that information is really destroyed. The
gist of our argument is this: {\it i)}Information loss indicates
indeterminism. {\it ii)}Indeterminism in itself is as time-symmetric as
determinism. {\it iii)}However, in a world in which entropy is increasing,
indeterminism indicates that causality is asymmetric: Low entropy events
determine high entropy events - but not {\it vice versa}. 

\section { Indeterminism Entails a Universal Time-Arrow }

There is a well-known yet crucial difference between the normal, entropy
increasing evolution, and the time-reversed, order increasing one. {\it The
latter, not the former, requires infinitely precise pre-arrangements of all the
system's elementary particles}. For a normal process, no special care is needed
to arrange its particles so as to increase entropy. Boltzmann's definition of
entropy, $S=k \ln W$ , indicates that there are numerous microscopic
arrangements that make disordered states but only few arrangements that make
ordered ones. Hence, nearly every initial arrangement will eventually give rise
to entropy increase. Consequently, a change in the initial arrangements can
hardly affect entropy increase (Fig. 2a). Not so with the time-reversed system:
The slightest change in the position or momentum of a single particle will
create a disturbance in the system's evolution that - given sufficiently many
interactions between the particles - will further increase as the system
evolves. Consequently, entropy will increase in the time-reversed system too
(Fig. 2b). As Yakir Aharonov vividly puts it, take out one worm from a dead
person's grave, and the time-reversed evolution will fail to bring him or her
back to life.

\begin{figure}
\centering
\includegraphics[scale=0.95]{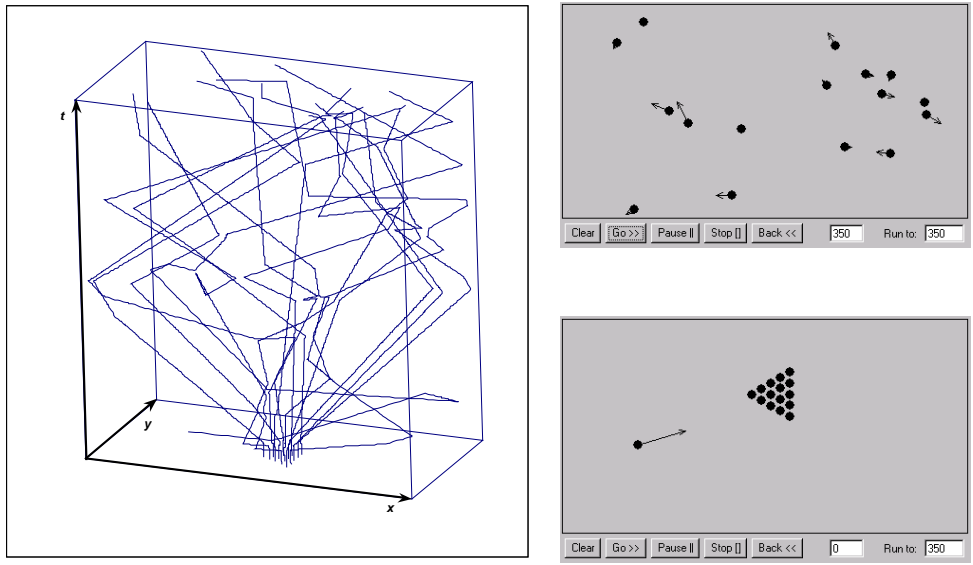}
\renewcommand{\thefigure}{1a}
\caption{A computer simulation of an entropy increasing process, with the
initial and final states (right) and the entire process using a spacetime
diagram (left). One billiard ball hits a group of ordered balls at rest,
dispersing them all over the table. After repeated collisions between the
balls, the energy and momentum of the first ball is nearly equally divided
between the balls.}
\end{figure}

\begin{figure}
\begin{center}
\includegraphics[scale=0.95]{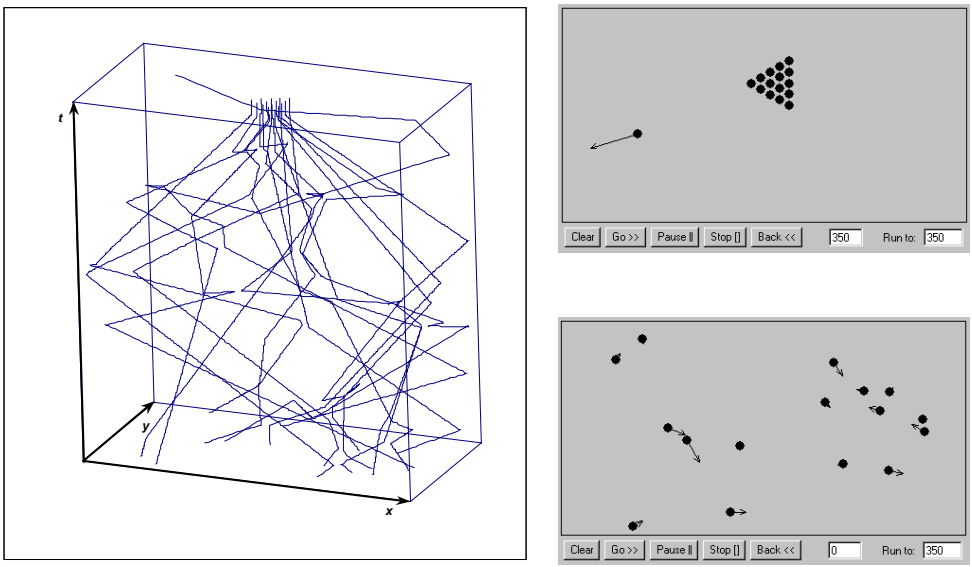}
\renewcommand{\thefigure}{1b}
\caption{The time-reversed process. All the momenta of the balls are reversed
at t=350. Eventually, the initial ordered group is re-formed, as at t=0, ejecting
back the first ball.}
\end{center}
\end{figure}

\begin{figure}
\begin{center}
\includegraphics{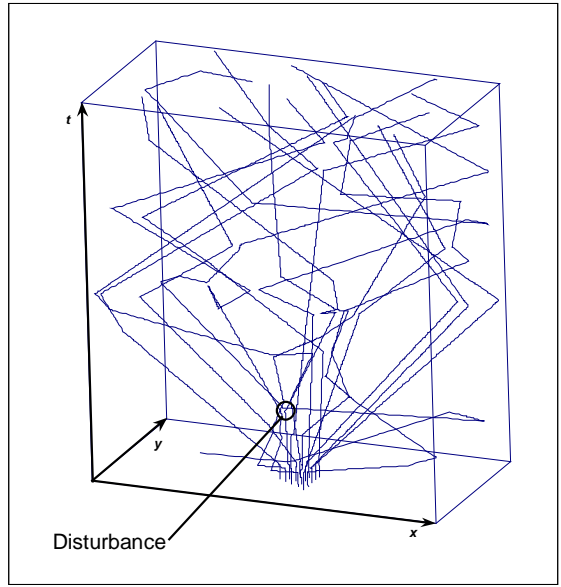}
\renewcommand{\thefigure}{2a}
\caption{The same simulation as in 1a, with a slight disturbance in the
trajectory of one ball (marked by the small circle). Entropy increase seems to
be indistinguishable from that of 1a.}
\end{center}
\end{figure}

\begin{figure}
\begin{center}
\includegraphics{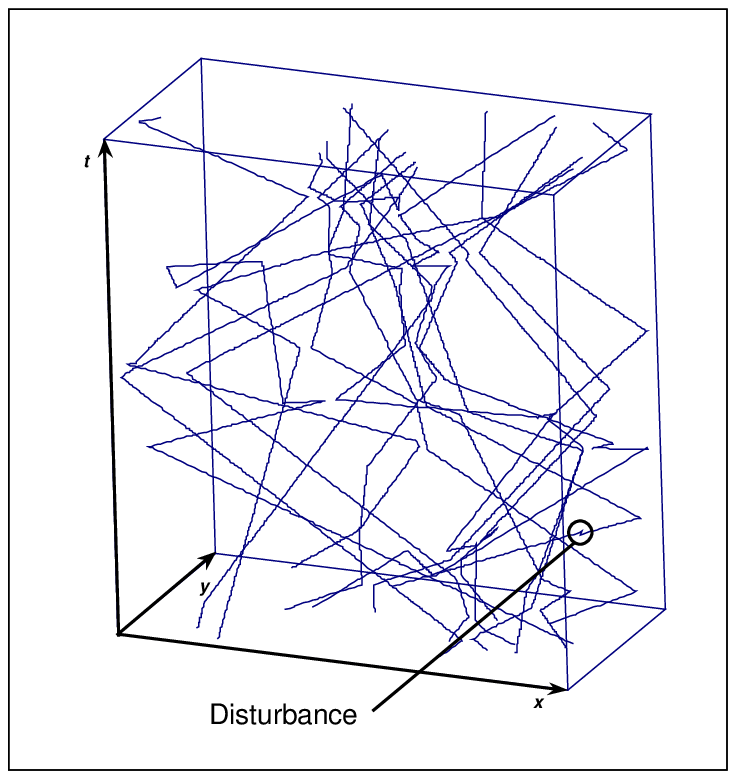}
\renewcommand{\thefigure}{2b}
\caption{The same computer simulation as in 1b, with a similar disturbance.
Here, the return to the ordered initial state fails.}
\end{center}
\end{figure}

This restriction is almost trivial, but its far-reaching consequences have not
been explored. Had physics been able to prove that determinism does not always
hold - that some processes are governed by fundamentally probabilistic laws -
it would follow that entropy {\it always } increases, regardless of the
system's initial conditions. An intrinsic time-arrow would then emerge in {\it
any } system, independent of the initial conditions. Rather, the emergent time
arrow would be congruent with that of the entire universe, of which closed
systems are supposed to be shielded.
 
\section {Quantum Mechanics does Not  Disprove Determinism}

At first sight, quantum mechanics seems to have disproved determinism long ago,
thereby giving an intrinsic time arrow. A closer consideration, however, shows
that QM has never ruled out the possibility that determinism still exists at
some unobservable level \cite{Eli2,Eli3}. 

To see this, let us examine an alleged demonstration quantum mechanical
indeterminism often used by Penrose \cite{Pen,Pen1}. A half-silvered mirror
splits the wave function of a single photon such that one half hits a detector
and the other half goes to the wall. In 50\% of the cases, the detector will
click. Suppose now that we time-reverse this process. The photon's wave
function will be split again by the beam splitter when going back, thereby
giving 50\% probability for not returning to the lamp but rather going to the
opposite wall. This, for Penrose, indicated time-asymmetry at the quantum
level. 

Recently, however, Penrose \cite{Haw2} conceded that this asymmetry may merely
be reflecting the asymmetry of the boundary conditions. In the normal
time-evolution, half of the wave function initially goes to the wall. Now this
half wave function is a real physical phenomenon: Reflected back by a mirror,
this half can be used to create interference effects (The Elitzur-Vaidman
bomb-testing experiment \cite{Eli4}, to which Penrose \cite{Pen1} gives a vivid
exposition, proves how real this "empty" half is.). Therefore, for a real time
reversal to take place, the wall's absorption of the half wave-function (i.e.,
its interaction-free measurement) must be reversed too. Once we took care to
make the time reversal thus complete, the photon would indeed return to the
lamp from which it was initially emitted. Penrose, of course, believes that
this would not happen, for he regards "collapse of the wave function" as a real
process. However, no experiment is known today that can favor this
interpretation over other, time-symmetric ones. For example, the "guide wave"
or the "many worlds" interpretations assume that some hidden variables, in the
form of empty waves or parallel universes, remain after the measurement,
preserving all the seemingly-lost information (see Unruh \cite{Unr} for another
objection to Penrose's argument).  

Quantum mechanics, therefore, just like classical physics, allows any process
to be time-reversed under the appropriate initial conditions.

\section {Hawking's Information-Loss Hypothesis}

It is black holes, however, that seem to provide what we are looking for. As
noted in the introduction, Hawking claims that all the in-falling matter's
information, save the conserved quantities M, Q, and J, is obliterated by black
hole evaporation. While a detailed review of the debate concerning Hawking's
hypothesis is beyond the scope of this paper, we shall briefly mention the
hypothesis and the counter arguments and then, without taking a stand, we shall
point out the bearing of this hypothesis on the question of time asymmetry. 

Due to the Hawking effect \cite{Haw4,Haw5}, black holes must eventually
evaporate. Since the resulting radiation comes from quantum vacuum fluctuations
at the black hole's horizon, it seems to be absolutely thermal, being unrelated
to, and preserving nothing of the black hole's content. This gives rise to
entropy that is not due to coarse graining, as in classical physics. Rather,
the entropy seems to be absolute. 

Naturally, Hawking's claim has raised several objections, yet none turned out
to be decisive. Especially the loss of unitarity seems disturbing. Among the
most radical attempts to preserve unitarity is t'Hooft's \cite{Hoo} S-Matrix
approach, which assumes that a pure state of a complete system would always go
to a pure state, hence no information is lost. But as t'Hooft himself admits,
the model is still hypothetical (see also Page \cite{Pag}). Other authors
pointed out that, for external observers, nothing really crosses a black hole's
horizon; the gravitational time-dilation red-shifts the fallings objects to the
point where they "freeze" on the horizon, never really appearing to cross it.
However, for a freely-falling observer, the horizon is devoid of physical
significance; it will be crossed soon, and no "frozen" object would appear. The
case is similar to that of black-hole formation: here too, the imploding matter
was believed to "freeze" before reaching the critical circumference. Yet
Finkelstein \cite{Fin} has shown how to reconcile the two accounts, that of the
external observer and that of the freely-falling one, into one, self-consistent
account. Arguments based on reference frames are therefore insufficient for
dismissing Hawking's paradox \cite{Bek}.

Preskill \cite{Pre}, initially Hawking's opponent, has thoroughly reviewed all
the proposals to avoid the information loss paradox and found all of them
deficient. "The information loss paradox," he concludes, "may be a genuine
failing of 20th Century physics, and a signal that we must recast the
foundations of our discipline."

We submit that, if Hawking's argument is sound, then its most immediate bearing
has gone unnoticed so far. It reveals a fundamental origin of time symmetry. 

Consider, then, the following thought experiment. Let a closed system undergo a
normal evolution, such that its entropy increases with time. Also, let the
system have enough mass and time to allow a black hole to form and evaporate.
Opening the system after sufficient time, we find that its entropy has
increased. This is not surprising: If Hawking's hypothesis is correct, the
particles into which the black hole has evaporated could not preserve the
positions and momenta of the objects swallowed earlier by the black hole.
Hence, the black hole has only added to the system's entropy, in a way similar
to that of the event in Fig. 2a where the causal chain was interfered with. 

Consider next the time-reversed system. Let there be a similar closed system,
with the positions and momenta of all its particles pre-correlated with great
precision such that its entropy would decrease with time. And here too, the
amount of matter and the time allocated to the system suffice for the formation
and evaporation of a black hole. Opening the system at the end of the
experiment, we find that the time-reversal has failed: Entropy has increased in
this case too. 

The reason is clear: The black hole's information destroying effect has ruined
the pre-arranged correlations with which the initial state has been prepared.
This parallels the simplified case shown in Fig. 2b, with the difference that
the failure of determinism entailed by the black hole effects not only one but
{\it numerous} particles.
Let us put this failure in physical terms. Preparing a black hole in a
time-reversed system amounts to preparing a {\it white hole}, i.e., a
singularity from which light as well as macroscopic objects are ejected. But
this is precisely what we cannot do. In order to create a white hole, numerous
thermal particles must be directed towards one point. This is the time-reversal
of the black hole's evaporation. With a sufficiently huge number of particles a
singularity will indeed form, which will later evaporate. However, when it
evaporates, then, by Hawking's hypothesis, it evaporates again into particles
whose spectrum is {\it thermal}; no objects with complex physical attributes
can emerge from it. Although Hawking seems to be unaware of it, his hypothesis
provides a perfect explanation for the absence of macroscopic white holes.

A normal system, then, increases its entropy when an information-erasing event
is formed within it, but so does a time-reversed system. The conclusion
therefore follows: {\it In any closed system that gives rise to an
information-erasing singularity, an intrinsic time-arrow emerges that
disregards the system's boundary conditions, but complies with the time-arrow
of the universe, of which closed systems are supposed to be shielded}.

\section { Summary}

In mainstream physics, the Second Law is not a real law but rather a
consequence of the universe's initial conditions; under special initial
conditions, the Second Law can be reversed. 

We have shown that this assertion is correct only under absolute determinism.
With even the slightest failure of determinism, an intrinsic arrow of time must
emerge in any closed system, regardless of its initial conditions, but with
perfect accordance with the time arrow of the entire universe, despite the
system's isolation. 

The implications of this inherent time arrow for the nature of time are very
far-reaching, and discussed in detail elsewhere \cite{Eli1}. If future events
are indeed devoid of causal efficacy on past events, then the awkward
possibility mentioned at the beginning of this paper, namely, that we live in
an entropy-decreasing universe while unable to observe it, would be ruled out
at last. But then, the very definition of time as a dimension in Minkowski's
4-geometry requires revision.
 
Our conclusions equally apply to all other theories that assume genuine
indeterminism. Thus, although proponents of the indeterministic interpretations
of quantum mechanics are generally unaware of it, their interpretations imply
that causality itself is the origin of time asymmetry. 

\section { Acknowledgments}

This work was supported in part by The Stewart and Judy Colton Foundation. It
is a pleasure to thank Yakir Aharonov for many illuminating discussions.

\bibliographystyle{unsrt}

\end{document}